# The Evolution of Teaching Methods of Students in Electronic Social Networks

Svitlana Lytvynova, Olga Pinchuk

Institute of Information Technologies and Learning Tools
of National Academy of Educational Sciences of Ukraine,
9 M. Berlyns'koho st., 04060, Kyiv, Ukraine
s_litvinova@i.ua, opinchuk100@gmail.com

**Abstract.** The use of electronic social networks (ESN) as a new teaching method that is chosen by modern teachers, who have sufficient professional, information and communication competence, passed several stages of formation and nowadays are widely used in the educational process in secondary schools.
The article deals with basic approaches to determining teaching methods in secondary schools through electronic social networks. Main approaches to the selection of teaching methods, particularly through the internal logical path of learning, are defined.
Authors demonstrate such teaching methods as: comparison, synthesis, summing up, specification and classification; analytical, synthetic and application of formative assessment. Examples and algorithm of use for each method are given in the article. It is established that the effectiveness of teaching methods depends not only on the same methods but on skills of a teacher to use the functionality of social networks and ICT.

**Key Terms.** TeachingMethodology, TeachingProcess, KnowledgeManagementMethodology, QualityAssuranceMethodology, StandardizationProcess, ICT

## 1 Introduction

The use of electronic social networks (ESN) as pedagogical innovation, new teaching method that is chosen by teachers who have sufficient professional information and communication competence, passed several stages of development and got acceptance of society. Detailed review [9; 10] of real-life experience and prospects of ESN in secondary education fulfilled by us, indicates that technological innovation in the course of its formation (Hype Cycle, Gartner) runs from the peak of popularity to the lowest point of frustration, and now can be at the stage of enlightenment (Fig. 1).

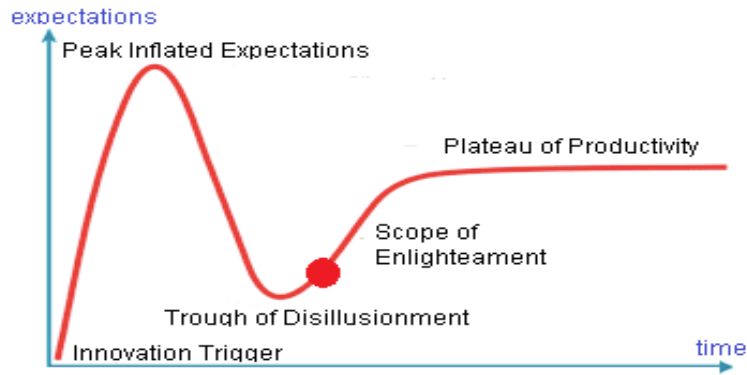

**Fig. 1.** Electronic social networks in the cycle beyond expectation

Mostly, teachers and parents of students are aware of the threats that are present in the ESN. Sense of social networks as an integral part of the modern world, awareness of their advantages, disadvantages and limitations are gradually formed in the pedagogical community. Ways of overcoming these defective features exist and are actively popularized. New positive examples of ESN implementing in the educational process appear.

With the introduction of new, modern information and communication technologies in the educational process of general education requirement in the selection of effective teaching methods arises.

## 2 Related Work

In the following study, we relied on the following definitions. The method (from the French. - Methode) – a way of research; method of theoretical research or practical implementation of something; method [12, p.364]. The method of teaching is a way of ordered and cooperative acts of teachers and students, aimed at solving problems of education, training and development in the learning process [5, p.129]. Teaching methods are necessary to ensure that a student has mastered the subject content of the study, learned to perform subject action, and actively operated ways of learning and working creatively [1]. Bilateral nature of teaching content should be displayed in teaching methods, which ones, on the one hand are various attempts to help a student in learning process, to contribute to its activization and on the other hand, they are ways for students to acquire competencies.

Knowledge and communication are inseparable in learning. Electronic social networks today are a popular means of establishing of social relationships and maintaining communication. In ESN communication of teachers and students goes through social objects (photos, videos, audio messages, presentations, surveys). Each of these objects can be an effective learning tool in a certain methodical system [2]. Social object can become an illustrative material for the lecture, a narration that is a result of individual work, problems setting to heuristic lesson during the stage of motivation of learning activities, means of control during updating of basic knowledge, etc. In this case, a teacher uses a variety of forms (collective, group or individual) with students. The choice of method of teaching, in our opinion, is the most significant.

There are several classifications of teaching methods, which are based on certain characteristics. The most successful, in our opinion, methods of teaching students of secondary schools while using various types of ESN (corporate, universal, theme, etc.), we consider those which guarantee logic of search, transmission and perception of educational material (data of teaching and cognitive content). These teaching methods are primarily associated with scientific methods, methods of knowledge, the logic of data assimilation, where analysis, synthesis, comparison, generalization, specification and selection of main points always present [5, p. 131].

## 3    Research Methods

The theoretical basis of the study are composed with conceptual provisions of methodology of scientific research. We used the theoretical conclusions of scientists on current state of education informatization (V.Bykov, R. Gurevich, M. Zhaldak, A. Gurzhiy etc.) and scientific and pedagogical basis of formation and use of information educational environment (V. Bykov, Yu. Zhuk V. Oleynik, Y. Polat etc.). Also, the most common in the general didactic classification of teaching methods of A. Aleksyuk, Yu Babanskii, G. Vaschenko M. Danilov, B. Yesipov, I. Lerner, M. Levin, M. Makhmutov, V. Okon', I. Podlasoho V. F. Palamarchuk, V.I. Palamarchuk, S.Petrovsky, A. Pinkevich, M. Skatkin, K. Sosnitsky A. Sohor', I. Kharlamov are included. As well as researches of creators of special learning methods, (those that are used in the study of certain subjects) are taken as a basis of research. Recent scientific results of practical use of ESN in general secondary education, including: the development of information and communication and social competencies in electronic social networks (V. Kovalenko, S. Lytvynova [2; 7; 8], O. Pinchuk [9; 10], O. Sokolyuk [14]); social networks as a means of independent activity of students (A. Slobodyanyk [13]), use of electronic social networks in work with supernormal pupil (N. Yaskova, A. Yatsyshyn [18]), electronic social networks as a component of modern social media (O. Konevschynska [6]). A number of theoretical methods: analysis of research problems in scientific publications; studying the experience of using electronic social networks in the learning process are used in the study. We use methods of comparative analysis in the study of the features of different teaching methods and varied features, which are the basis for classification.

## 4    Research Findings

We believe that educational outcomes that can be achieved in the educational process using electronic social networks, primarily detected in ability of students to work individually or collectively, applying tools of computer networks, information and resources of information and educational environment, obtain and responsibly use data to solve educational problems, communication, acquire new knowledge.

Analysis of educational and cognitive activity of a student in information and educational environment, comparing it with categories of cognitive processes of Modified Bloom's Taxonomy and description of categories of Bloom's Digital Taxonomy, allowed to draw conclusions about the powerful impact of ICT on basic processes in the educational system: transferability of learning, acquiring skills, record of achievement, assessment of education quality, creation of positive motivation and promotion of autonomy in teaching and learning of [14, p. 33].

We agree with the opinion of a researcher that ESN provide new tools of educational processes. Systematization of types of pupils' educational and cognitive activity, ICT tools and network services in accordance with the categories of cognitive processes create an objective basis for updating approach to use teaching methods that are grouped by the nature of cognitive activity of students (Table 1).

**Table 1.** Methods of teaching by the nature of cognitive activity.

|  | Cognitive activity | Learning methods | Advantages |
|---|---|---|---|
| Performing | Awareness, memorising, reproduction of information | Explanatory illustrative | Systematicity, subsequence, time management |
| Performing | Reproduction of methods of activities defined by the algorithm | Reproductive | Large amount of information in a short time |
| Transitional | Creative teaching and learning activities on the basis of apperception and reliance on previous experience | Problem-based learning | Formation of student's abilities to solve problem tasks during the process of observations of teacher's cogitation |
| Creative | Partly independent obtaining of knowledge under guidance of a teacher | Partly research | The increase of cognitive independence |
| Creative | Creative application of knowledge, mastering methods of scientific cognition | Explorative | Activates cognitive activity, acquaints students with the stages of research |

Practical experience with ESN in education indicates the appropriateness of their use to develop independence of thinking schoolchildren, forming a critical attitude to educational information, and thus increases the effectiveness of partially-search methods.

However, the most successful in terms of using various types of ESN (corporate, multiuse, theme, etc.) we consider approaches to study methods of teaching students based on internal logical way of learning. Separately we consider the following methods: comparison, synthesis, summing up, specification, classification, analytical, synthetic, inductive, deductive, intermediate control - and detail some of them.

*Comparison* is a method of learning to identify similarities and differences between objects or phenomena.

The method of comparison involves the following: definition of objects of comparison; identification of the major features; comparison; finding similarities or differences; symbolic presentation of results of comparison (tabling, designing a plan, scheme or model). This method is used for separation of essential and inessential characteristics in comparable objects.

The algorithm of application:
- ensure that compares homogeneous phenomenon;
- determine the full amount of similarity signs;
- establish the total number of differences signs;
- conclude about common and different [15, p.9].

*Generalization* is a method of learning that is based on transition from the individual to more general knowledge, from the general one of a certain level to a general higher level, abstraction and finding common features that are peculiar subject's sciences (or discipline).

It is used when students must learn both how to classify as educational material at different stages of a lesson and objects, phenomena, types, groups, etc. Such actions are specific to generalization: selection of typical facts, finding essential; comparison; primary pins. Their theoretical interpretation; analysis of events dialectics; symbolic presentation of results summary (formulas, graphical models, tendencies, etc.).

The algorithm of application:
- compare objects and identify a total number of similarity and difference features;
- identify significant and insignificant features;
- combine similar objects by essential features in a single conjunction;
- conclude about objects by essential features.

Let us consider this method on example of the tasks, which were offered to students in a group class in ESN during completion of a theme "Quadrilaterals" (Grade 8).

Technology: Students are offered the table that contains a comprehensive list of properties of figures being studied: a random parallelogram, rectangle, and rhombus, square, trapezoid. The task: to present an array of quadrangles using Euler - Venn diagram. Stored pictures are placed on the student's "wall". Teacher and students commented on the results in a common chat or private messages.

Summing up - learning method, which involves the specification of the object of knowledge, sharing of data into logical parts, group and comparing them, separating the primary from the secondary.

This method is characterized by: the action of finding keywords, concepts, semantic bases; grouping the material; conclusion on the subject of knowledge, symbolic design (plan, scheme, reference outline algorithm, the title). The method is often used for theoretical summing up, for getting rid of unnecessary information that students' book contain. Such approach appears on different stages of a lesson: setting up a task, recitation, revision and especially on the stage of learning new material.
The algorithm of application:
- reduce an object into its component parts;
- compare components of the object;
- identify essential (the main) features of the object.

Let us consider the complex of learning methods in social networks on example of social object "presentation" (Fig. 2).

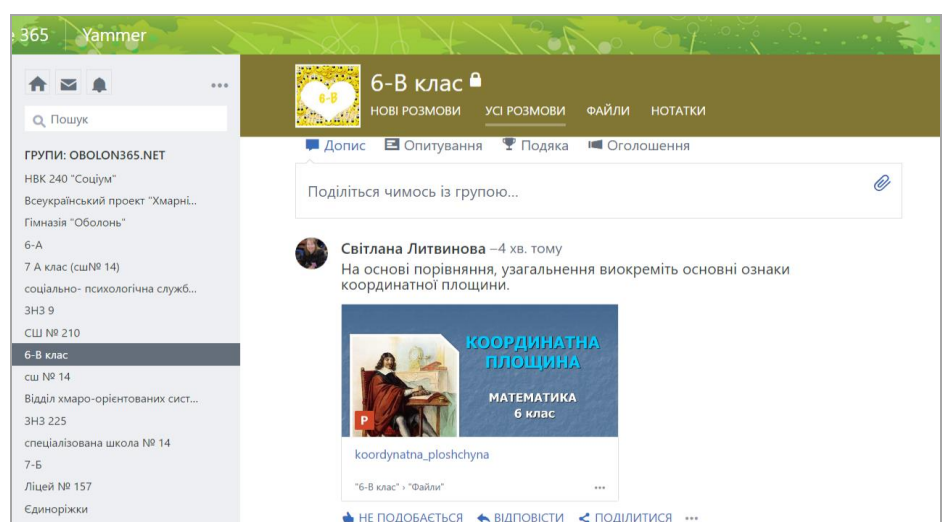

**Fig. 2.** A sample of complex task

Technology: Students study individually a presentation created by a teacher, come to know a new material, put questions in chat, and express their views on the essential features of the Cartesian plane on the "wall". The teacher answers students after receiving offers to participate in the blitz-survey to determine the Cartesian coordinate system. During real-life communication teacher summarizes the opinion of students, formulates definitions, and comments on the practical tasks depicted in the presentation.

*Specification* is a teaching method that provides the transition from the abstract to the concrete.

The method is characterized by the following: ascendancy from the abstract to the concrete; symbolic presentation of specification results (examples, problems, diagrams, models, etc.). It is used to specify the terms or existence of the phenomenon to gain theoretical knowledge with case studies.

Social networks contain plenty of links to videos and animated stories (Fig. 3-4), which can be subject to observation and "a starting point" for the transition from direct experience to the heart of the observed phenomena.

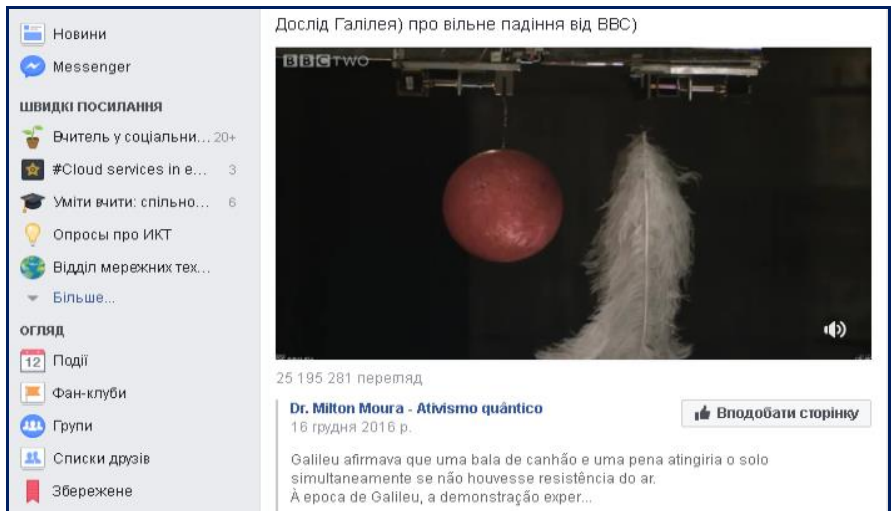

**Fig.3.** Galileo's experiment with cannon-shot and feathers carried out in modern laboratories

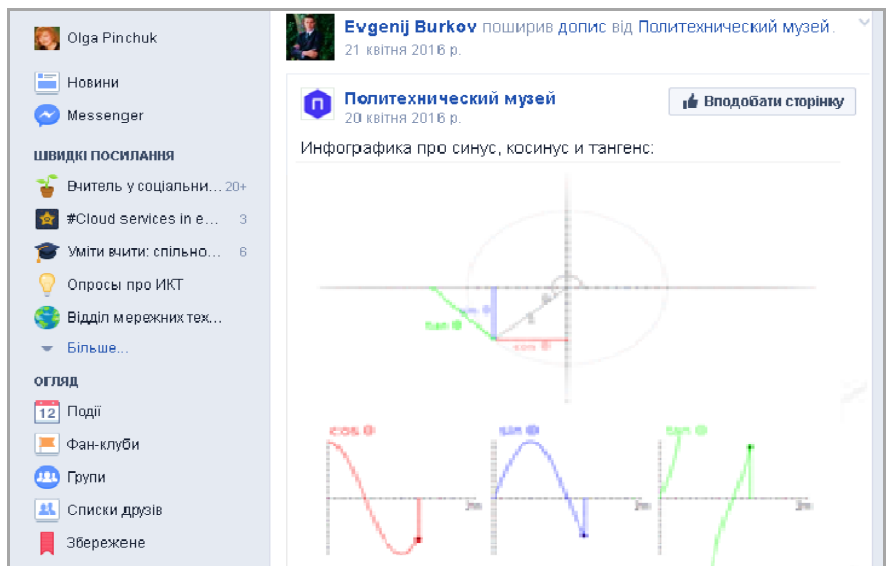

**Fig. 4**. Plotting of goniometric functions

*Classification* is a learning method, which involves the process of searching and finding the essential and common characteristics, elements and links for some of the objects that form the basis for group-reduced distribution [15, p. 9].

Let us consider the use of classification on example of social object "photo" (Fig. 5)

**Fig. 5.** Sample of classification method use

Technology: students receive a scheme from a teacher for analysis and suggested to classify geometric shapes independently; students can specify the objects in the diagram and ask questions via instant messages. Then, students are asked to answer questions blitz - survey (number of figures obtained in each classification group) to find out the problems they have in understanding of educational material.

*Analytical.* This method consists in selection of specific characteristics of the phenomenon and its expansion to the elements (Fig. 6).

**Fig. 6.** Sample of the task with use of analytical method

It has the following components: perception of meaningful information, isolating essential features and relationships, division finding the original elements and structural units; understanding links, their synthesis. Parts the whole are characterized by comparison, logic synthesis and other techniques. The method of analysis is often used on the initial (empirical) stage of knowledge [3].

The algorithm of application:
- expand object into its component parts;
- identify essential features of the object separately;
- examine separately each object sign as part of a whole;
- connect sections of an object into a single unit;
- summarize information about an object by essential characteristics [15, p.10].

The method lies in combining previously separate parts by analysing the elements or properties of the object together. It provides specific knowledge through the unity of diverse and mostly carried out on a theoretical level of knowledge [3]. Effective use of analysis and synthesis as learning methods is provided their interaction, indicating the term "an analytical and a synthetic method".

An important in learning is an application of critical thinking to construct reasonable conclusions. Inductive and deductive learning methods could be the basis for developing skills of students to draw conclusions about the object. These methods are actively used by teachers in a network of educational projects.

*Formative assessment as* a training method is aimed at monitoring or analysis of current students' knowledge of a particular topic of study. The most common are electronic survey and testing. For example, in a group of social network a public survey can be held or with the aim of assessment answers can be collected using Google Forms or Microsoft Forms links.

## 5  Discussion

During the research a discussion on the typing didactic potential of different social networks came up.

We came to a decision that the didactic potential of social networks for implementation of teaching methods discussed by us have some differences. However, the teacher has his choice of social network based primarily on the range of his own competence and preference. A teacher uses a ESN, encourages students to cooperate actively on it, chooses such learning methods that could be implemented exactly in this network (Fig. 7-8). For instance, one teacher manages the independent extracurricular work of students using teachers' blog on Blogspot, and the other one uses the potential of organization and management on Facebook.

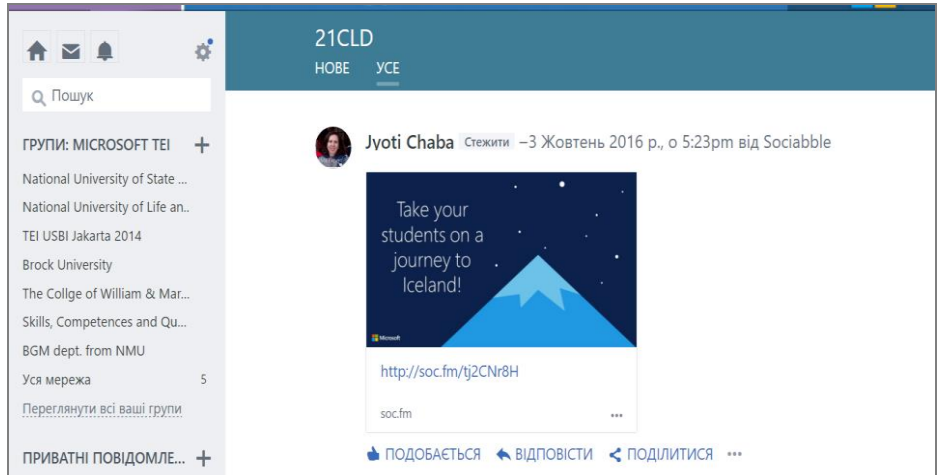

**Fig. 7.** Virtual tours

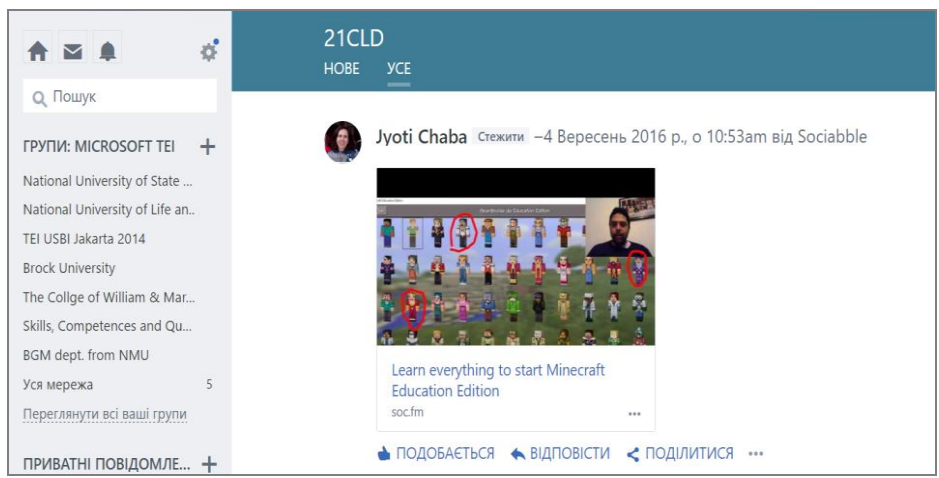

**Fig. 8.** Gamification

## 6   Conclusions and Prospects of Further Research

Together with the term "learning methods", steady combination: "an art of a teacher", "teaching tool", "multi qualitative phenomenon" are used in scientific and methodological sources. Teachers integrate different methods in practice creatively. Methods of teaching according to logic of educational information assimilation reflect nature and logic of disclosing of educational material content. Their use actively influence development of students' abstract thinking, formation of their concept system and cause-and-effect relationship.

The choice of learning methods depends on the objectives and content of the educational material of each lesson; characteristics of relevant scientific field methodology, characteristics of specific teaching methods of an academic discipline; time spent on the study of a material; age peculiarities of students; their skills; equipment and material procurement of educational process, including equipment and others.

The practical experience of authors gives ground to believe that the use of ESN helps to create a situation of interest in learning when teaching a particular educational material. The development of students' interest is an effective means of learning intensification that promotes better learning, encourages independent learning activities.

Task performance in cooperation with the help of information and communication networks encourages the duty and responsibility of students, and thus have a positive educational effect.

The effectiveness of methods depends on not only the methods themselves, but the skill of the teacher to use the functionality of social networks and ICT.